\documentclass[aps,prl,showpacs,notitlepage,singlecolumn,superscriptaddress,11pt,floatfix,]{revtex4-2}

\usepackage{mathtools}
\usepackage{multirow}
\usepackage{xcolor}
\usepackage{bm}
\usepackage{amsfonts,amssymb,amsmath}
\usepackage{graphicx,dcolumn,bm,xcolor,braket,slashed}
\usepackage{times} 
\usepackage{comment}
\usepackage{array}
\usepackage{textcomp}
\usepackage[normalem]{ulem}
\usepackage{dsfont}
\usepackage[colorlinks=true,citecolor=blue,urlcolor=blue]{hyperref}

\usepackage[title,toc,titletoc,page]{appendix}

\newcommand{\be}{\begin{eqnarray}}
\newcommand{\ee}{\end{eqnarray}}
\newcommand{\bbm}{\begin{bmatrix}}
\newcommand{\ebm}{\end{bmatrix}}
\newcommand{\bpm}{\begin{pmatrix}}
\newcommand{\epm}{\end{pmatrix}}

\begin{document}

\title{Color and Transparency from Quantum Geometry}
\maketitle

\vspace{-25pt} 
\begin{center}
{Chang-geun Oh,$^{1,*}$ and Sun-Woo Kim,$^{2,\dagger}$
}
\vspace{5pt}

\emph{$^{1}$Department of Applied Physics, The University of Tokyo, Tokyo 113-8656, Japan}\\
\emph{$^{2}$Department of Materials Science and Metallurgy, University of Cambridge,\\ 27 Charles Babbage Road, Cambridge CB3 0FS, United Kingdom}\\

$^{*}$ \textup{\href{mailto:cg.oh.0404@gmail.com}{cg.oh.0404@gmail.com};}
$^{\dagger}$ \textup{\href{mailto:swk38@cam.ac.uk}{swk38@cam.ac.uk}}\end{center}


\section{abstract}
The optical properties of solids are governed not only by their energy band dispersions but also by the quantum geometry of Bloch states. While the role of energy bands in determining the perceived optical appearance of materials, such as color and transparency, is well established, the influence of quantum geometry remains elusive.
Here, we demonstrate that the color and transparency of materials can be direct manifestations of their underlying quantum geometry. 
To illustrate this principle, we employ quadratic band-touching models that allow us to tune only the geometric properties of Bloch states, while keeping the energy dispersion fixed. This decoupling reveals that modifying the wavefunction texture alone can lead to dramatic changes in the optical conductivity and, consequently, in the reflectance spectrum of the material.
This results in distinct and controllable changes in perceived color. Similarly, we show that quantum geometry can govern the transparency of two-dimensional materials.
Our findings demonstrate how quantum geometry shapes the visual appearance of materials, opening new avenues for tailoring color and transparency beyond traditional band structure design. This establishes quantum geometric engineering as a novel approach for manipulating materials with customized optical functionalities.

\section{Introduction}
Quantum geometry has emerged as a central concept in modern condensed matter physics, providing a unifying framework to understand a vast range of quantum phenomena\,\cite{peotta2015superfluidity, torma2022superconductivity, ozawa2021relations, yu2024quantum,torma2023essay}. The geometry of quantum states is described by the quantum geometric tensor (QGT), whose symmetric real part is the quantum metric, and whose antisymmetric imaginary part is the Berry curvature\,\cite{provost1980riemannian, ma2010abelian, berry1989quantum}.
The role of the Berry curvature is well-established as the foundation of topological physics. 
Its integral over the Brillouin zone defines topological invariants, which characterizes phases of matter such as Chern insulators and the quantum Hall effect\,\cite{nagaosa2010anomalous, Xiao_2010_RMP}.
In contrast, the physical consequences of the quantum metric have only recently begun to be explored in depth such as electron-disorder scattering\,\cite{oh2024thermoelectric}, superfluid weight in superconductors\,\cite{torma2022superconductivity, peotta2015superfluidity, liang2017band}, anomalous Landau level spreading\,\cite{rhim2020quantum,hwang2021geometric}, superfluid weight of exciton condensate\,\cite{verma2024geometric,HuPRB2022}, the quantum Hall effect in bilayer graphene\,\cite{oh2024revisiting}, fidelity susceptibility near topological phase transitions\,\cite{panahiyan2020fidelity}, and bulk-interface correspondence in singular flat band systems\,\cite{oh2022bulk,kim2023general}. 

The influence of quantum geometry on the optical properties of solids is particularly profound, as the optical conductivity is directly related to the QGT\,\cite{cook2017design, de2017quantized, bhalla2022resonant,ghosh2024probing, ezawa2024analytic,ahn2022riemannian, chen2025dielectric,ezawa2025quantum}. In many systems, however, the optical response is a convolution of both the band dispersion and the quantum geometry, making it difficult to isolate the purely geometric contributions. A significant breakthrough in this context was the recent discovery of a universal, mass-invariant optical conductivity in two-dimensional (2D) quadratic band-touching systems\,\cite{oh2025universal}. This revealed that under specific conditions, the interband optical conductivity can be determined solely by quantum geometry, independent of material-specific band structure details, thereby providing a direct probe into the geometric nature of Bloch states.

Building on this understanding, we turn to the most fundamental and intuitive optical properties of materials: color and transparency. The color perceived by the human eye depends on the frequency-dependent reflectance spectrum within the visible range for opaque materials, while transparency is characterized by the ability to transmit visible light, a property common to most two-dimensional materials. 
Traditionally, these optical responses are attributed to the energy band levels of materials, the eigenvalues of the electronic band structure, which have been the primary focus in tailoring color and other optical properties\,\cite{nassau2001physics,fox2010optical,Kim2023_color}. This raises a fascinating question: Can the color or transparency of a material be manipulated not by changing its energy band dispersions, but by engineering the more subtle geometric properties of its quantum states? An affirmative answer would establish a new paradigm for designing materials with tunable color  or transparency.

In this work, we demonstrate that the color and transparency of materials can indeed be direct manifestations of their underlying quantum geometry. We introduce a three-dimensional (3D) model system where the geometric parameters of Bloch wavefunctions can be continuously tuned, while the energy dispersion is held fixed. This purely geometric tuning directly engineers the interband optical conductivity and, consequently, the reflectance spectrum of the material, leading to visible and controllable changes in color. We further analyze a 2D counterpart, illustrating that transparency can be governed by quantum geometry. Taken together, our findings introduce ``quantum geometry engineering" as a new paradigm for designing materials with bespoke optical functionalities, from color to transparency.

\section{Results}
\subsection{Quantum Geometry and Optical Conductivity}
\label{sec:qgeometry}
We begin by introducing the optical conductivity formula that is a central quantity used in this work to calculate color and transparency. The optical conductivity is given by the Kubo formula, 
\begin{align}
\sigma_{ij}(\omega) = \frac{e^2}{\hbar}\int \frac{d^dk}{(2\pi)^d} \sum_{n,m} F_{nm}(\bm{k}) \frac{i \epsilon_{mn}(\bm{k}) A^{i}_{nm}(\bm{k}) A^{j}_{mn}(\bm{k})}{\epsilon_{nm}(\bm{k}) + \hbar \omega + i \eta}, \label{eq:opt_cond}
\end{align}
where $F_{nm}(\bm{k})= f(\epsilon_n(\bm{k}))- f(\epsilon_m(\bm{k}))$, $f(\epsilon)=1/[1+e^{(\epsilon-\mu)/{kT}}]$ is 
the Fermi distribution function, $\epsilon_n(\bm{k})$ is the $n$-th band energy, $\epsilon_{nm}(\bm{k})=\epsilon_n(\bm{k})-\epsilon_m(\bm{k})$, $\mu$ is the chemical potential, and $\eta$ is a broadening parameter (inverse lifetime).
The Berry connection $A^{j}_{nm}(\bm{k})$ is 
\begin{align}
A^{j}_{nm}(\bm{k}) = i \langle u_n(\bm{k}) | \partial_{k_{j}} | u_m(\bm{k}) \rangle,
\end{align}
where \( | u_n(\bm{k})\rangle \) is the cell-periodic Bloch state.
Hence, optical conductivity intrinsically encodes quantum geometry through the Berry connection, showing the link between quantum geometry and the interaction of materials with light and, consequently, their visual appearance. In three dimensions, this geometry-dependent optical conductivity shapes the reflectance spectrum that determines perceived color, while in two dimensions, it governs absorbance that dictates transparency.
To demonstrate this principle concretely, in the following sections, we will analyze model systems where the quantum geometry can be tuned independently of the energy band dispersion.

\subsection{Color shaped by quantum geometry in 3D}

\subsubsection{Optical Reflectance in 3D Materials}
For the convenience of the reader, we review the connection between the macroscopic optical properties of a material and its microscopic electronic structure. The propagation of an electromagnetic wave in a medium is governed by Maxwell's equations. For a non-magnetic, homogeneous medium, these equations lead to the wave equation for the electric field $\bm{E}$:
\begin{align}
-\nabla^2 \bm{E}(\bm{r}, t) = \frac{1}{c^2} \frac{\partial^2 \bm{E}(\bm{r}, t)}{\partial t^2} + \mu_0 \frac{\partial \bm{J}(\bm{r}, t)}{\partial t},
\end{align}
where $\bm{J}$ is the induced current density. Assuming an Ohmic response $\bm{J} = \sigma(\omega) \bm{E}$ for a monochromatic plane wave $\bm{E}(\bm{r}, t) = \bm{E}_0 e^{i(\bm{k}_{light} \cdot \bm{r} - \omega t)}$, we arrive at:
\begin{align}
\nabla^2 \bm{E} = -\frac{\omega^2}{c^2} \left(1 + i\frac{\sigma(\omega)}{\epsilon_0 \omega}\right) \bm{E} \equiv -\frac{\omega^2}{c^2} \epsilon(\omega) \bm{E}. \label{Maxwell_eq_revised}
\end{align}
Here, we have introduced the complex, frequency-dependent dielectric function $\epsilon(\omega)$. This fundamental relation bridges the quantum mechanical optical conductivity $\sigma(\omega)$ from Eq.~\eqref{eq:opt_cond} to the dielectric response of materials.

The complex dielectric function is also expressed in terms of the complex refractive index $\tilde{n} = n + i\kappa$:
\begin{align}
\epsilon(\omega) = \tilde{n}^2(\omega),
\end{align}
where $n$ is the refractive index and $\kappa$ is the extinction coefficient.

For an electromagnetic wave propagating along the $z$-direction, the wave vector of the light, $k_{light}$, becomes complex:
\begin{align}
k_{light} = \frac{\omega}{c}\tilde{n} = \frac{\omega}{c}(n + i\kappa).
\end{align}
The plane wave solution inside the medium thus takes the form:
\begin{align}
\bm{E}(z,t) = \bm{E}_0 e^{-\kappa\omega z/c} e^{i(n\omega z/c - \omega t)}.
\end{align}
The term $e^{-\kappa\omega z/c}$ describes the exponential decay of the wave amplitude as it propagates through the medium, a phenomenon known as absorption. The term $e^{i(n\omega z/c - \omega t)}$ indicates that the wave propagates with a phase velocity $v_p = c/n$.

To connect these optical constants to the optical conductivity, we decompose $\epsilon(\omega)$ and $\sigma(\omega)$ into their real and imaginary parts:
\begin{align}
  \epsilon(\omega) = \epsilon_1 + i\epsilon_2,~~~~
  \sigma(\omega) = \sigma_1 + i\sigma_2.
\end{align}
The real part of the optical conductivity, $\sigma_1$, represents dissipation (absorption), while the imaginary part, $\sigma_2$, relates to the out-of-phase response. These two components are linked by the Kramers-Kronig relation:
\begin{align}
\sigma_2(\omega)=-\frac{2\omega}{\pi}\mathcal{P}\int_0^\infty d\omega' \frac{\sigma_1(\omega')}{\omega'^2-\omega^2}. \label{eq:KK_rel}
\end{align}
From the definition in Eq.~\eqref{Maxwell_eq_revised}, we find the explicit relations:
\begin{align}
\epsilon_1 = 1 - \frac{\sigma_2}{\epsilon_0\omega}, \quad \epsilon_2 = \frac{\sigma_1}{\epsilon_0\omega}.
\end{align}
Furthermore, by equating $\tilde{n}^2 = (n+i\kappa)^2$ with $\epsilon_1 + i\epsilon_2$, we obtain:
\begin{align}
  \epsilon_1 = n^2 - \kappa^2,~~~~
  \epsilon_2 = 2n\kappa.
\end{align}
This set of equations demonstrates that the optical constants $(n, \kappa)$ are fully determined by the real and imaginary parts of the optical conductivity $(\sigma_1, \sigma_2)$.

Finally, we consider the observable quantities for a semi-infinite medium with a planar interface at $z=0$, under normal incidence from vacuum. The reflectance $R$ (the fraction of reflected power) is given by:
\begin{align}
  R &= \left| \frac{1-\tilde{n}}{1+\tilde{n}} \right|^2 = \frac{(n-1)^2 + \kappa^2}{(n+1)^2 + \kappa^2}.
\end{align}
The transmittance $T_{int}$ (the fraction of power transmitted \textit{across the interface}) is given by $T_{int} = 1 - R$. For an absorbing medium ($\kappa > 0$), this transmitted light is eventually absorbed within the material. The perceived color of a thick, opaque material is thus primarily determined by the spectral features of its reflectance $R(\omega)$.

In summary, the quantum geometry of the electronic wavefunctions, as captured by the Berry connection in Eq.~\eqref{eq:opt_cond}, dictates the optical conductivity $\sigma(\omega)$. This, in turn, governs the optical constants $(n, \kappa)$ and ultimately determines the reflectance spectrum $R(\omega)$, which is perceived by an observer as the color of materials.

\begin{figure*}[t]
\includegraphics[width=140mm]{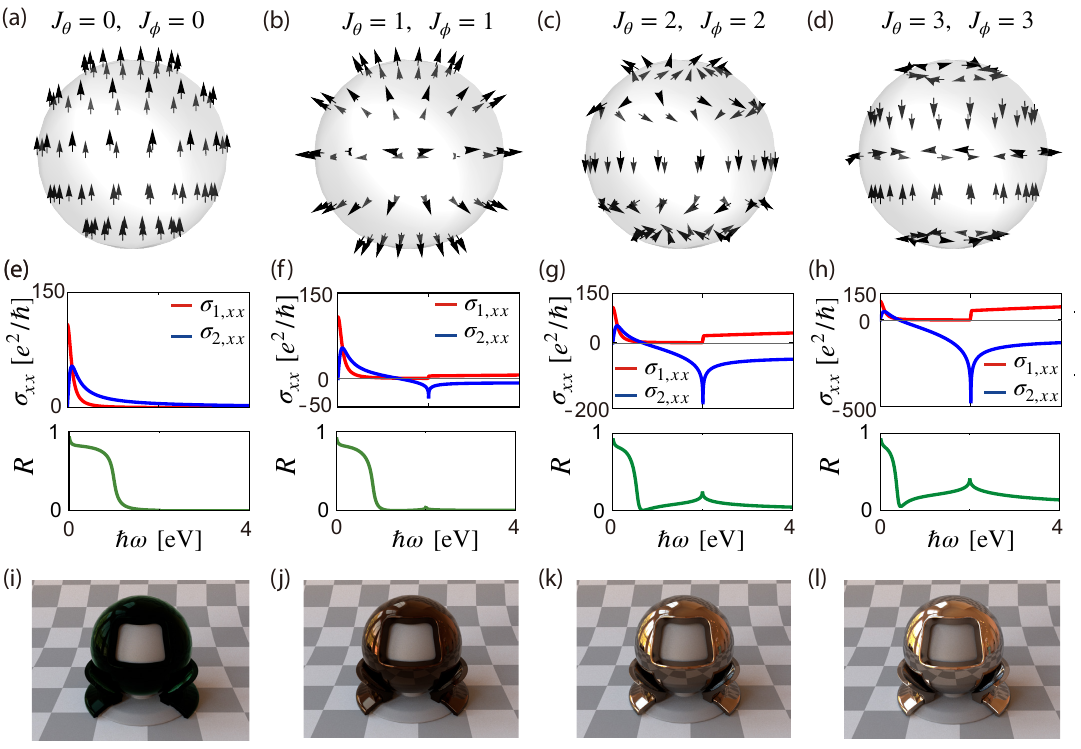} 
\caption{\label{fig1}
(a-d) Pseudospin textures of the model in Eq.~(\ref{eq:Ham3D}) for ($J_\theta,J_\phi$) = (0,0), (1,1), (2,2) and (3,3), respectively. 
(e-h)  Top panels: frequency dependence of the optical conductivity $\sigma(\omega)$ for the same ($J_\theta,J_\phi$) values; the red and blue curves denote the real and imaginary parts, respectively. Bottom panels: corresponding reflectance $R(\omega)$ (green curves).
(i-l) Photorealistic visualizations of the perceived material colors corresponding to the reflectance spectra shown in (e-h).
The photorealistic renderings are generated using the Mitsuba 3 renderer\,\cite{mitsuba3}, where the material is simulated on a standard test model that surrounds a central grey reference area. This setup enables consistent visual assessment of color and appearance. An ideal bulk system with a clean surface is assumed in all cases.
In the calculations, we set $m=0.1 m_e$, $m_e=0.511~\text{MeV}$, $\mu=1~\text{eV}$, and $\eta=0.1~\text{eV}$.
}
\end{figure*}

\subsubsection{3D Quadratic Band Touching Model}

To illustrate the connection between quantum geometry and the color of materials, we consider a toy-model Hamiltonian that describes a quadratic band touching point in three dimensions:
\begin{align}
    H(\bm{k}) = \frac{(\hbar k)^2}{2m} \bm{d}(\theta, \phi) \cdot \bm{\sigma},
\end{align}
where $\bm{\sigma}=(\sigma_x, \sigma_y, \sigma_z)$ is the vector of Pauli matrices and the direction vector $\bm{d}$ is defined on the unit sphere using spherical coordinates $k=\sqrt{k_x^2+k_y^2+k_z^2}$, $\theta=\arctan(\sqrt{k_x^2+k_y^2}/k_z)$, and $\phi=\arctan(k_y/k_x)$:
\begin{align}
    \bm{d}(\theta, \phi) = \bigg(\sin(J_\theta \theta)\cos(J_\phi \phi), \sin(J_\theta \theta)\sin(J_\phi \phi), \cos(J_\theta\theta)\bigg).\label{eq:Ham3D}
\end{align}
The dimensionless integers $(J_\theta, J_\phi)$ are geometric parameters that characterize the winding of the pseudospin texture in momentum space, as shown in Fig.~1(a). Here, the pseudospin is defined as $\bm{s}(\bm{k})=\bm{d}(\theta,\phi)$, which is scale-invariant with respect to $k$. The energy eigenvalues of this Hamiltonian are given by
\begin{align}
    \epsilon_\pm (\bm{k}) = \pm \frac{\hbar^2k^2}{2m}.
\end{align}
Crucially, the energy dispersion is independent of $J_\theta$ and $J_\phi$. However, these integers fundamentally alter the geometry of the Bloch wavefunctions. As we will show, this change in quantum geometry directly manifests in the optical response of materials and, consequently, their perceived color.

To quantify this effect, we calculate the optical conductivity of the model. The detailed derivation is provided in the Appendix. The diagonal component of the conductivity tensor is $\sigma_{jj}(\omega) = \sigma_{1,jj}(\omega) + i \sigma_{2,jj}(\omega)$, where the real and imaginary parts each contain both intraband and interband contributions.

The real part of the optical conductivity is given by:
\begin{align}
    \sigma_{1,jj}(\omega) = \underbrace{\frac{e^2}{\hbar}\frac{k_F^3}{6\pi^2}\frac{\eta}{(\hbar \omega)^2 + \eta^2}}_{\text{Intraband (Drude)}} + \underbrace{\frac{e^2}{\hbar}\frac{J^j_{J_\theta,J_\phi}\sqrt{m\omega/\hbar}}{32\pi^2}\Theta(\hbar\omega-2\mu)}_{\text{Interband}}.
\end{align}
The imaginary part $\sigma_{2,jj}(\omega)$ can be found via the Kramers-Kronig relation in Eq.~\eqref{eq:KK_rel}, yielding
\begin{align}
    &\sigma_{2,jj}(\omega) = \frac{e^2}{\hbar}\bigg(\frac{k_F^3}{6\pi^2}\frac{\hbar \omega}{(\hbar \omega)^2 + \eta^2} -\frac{J^j_{J_\theta,J_\phi}\sqrt{m\omega/\hbar}}{32\pi^2}\ln \bigg|\frac{(k_F+\sqrt{m\omega/\hbar})}{(k_F-\sqrt{m\omega/\hbar})}\bigg|\bigg).
\end{align}
Here, the chemical potential $\mu = (\hbar k_F)^2/(2m)$ defines the Fermi momentum $k_F$. The first term in $\sigma_{1,jj}$ is the standard Drude peak for intraband transitions, broadened by a scattering rate $\eta$. The second term describes interband transitions, which are only possible when the photon energy $\hbar\omega$ exceeds the vertical transition energy at the Fermi surface, $2\mu$, as enforced by the Heaviside step function $\Theta(\cdot)$. The interband contribution is derived in the small $\eta$ limit.

The key result is that the strength of the interband absorption is directly proportional to the geometric factor $J^j_{J_\theta,J_\phi}$, defined as:
\begin{align}
    J^j_{J_\theta,J_\phi} = \frac{1}{2}(J_\theta^2 C_\theta^j + J_\phi^2 C_\phi^j),
\end{align}
with the coefficients
\begin{align}
    C^i_\theta = \frac{\pi(2+6\delta_{iz})}{3}, \quad C^i_\phi(J_\theta) =(1-\delta_{iz}) \pi \int_0^\pi d\theta\frac{\sin^2(J_\theta \theta)}{\sin (\theta)}.
\end{align}
This explicitly demonstrates that the quantum geometry, parameterized by $(J_\theta, J_\phi)$, governs the optical response.
Note that the interband optical conductivity depends on the mass in the 3D quadratic band-touching model, whereas the 2D quadratic band-touching model exhibits mass-invariant interband optical conductivity\,\cite{oh2025universal}, as shown in Eq.~(\ref{eq:cond_2D_model}).

The consequences of this finding are illustrated in Fig.~\ref{fig1}.
Figures~\ref{fig1}(a-d) display the pseudospin textures for different integer pairs $(J_\theta, J_\phi) = (0,0), (1,1), (2,2), (3,3)$. These plots visualize how, despite sharing the same quadratic band dispersion, the systems possess distinct quantum geometries.
The resulting impact on their optical properties is shown in Figs.~\ref{fig1}(e-h), which plot the real and imaginary parts of the optical conductivity, $\sigma_{xx}(\omega)$ (upper panels), and the corresponding normal-incidence reflectance, $R(\omega)$ (lower panels). Variations in the geometric factor $J^x_{J_\theta,J_\phi}$ lead to significant changes in the interband conductivity, particularly creating prominent peaks, which, in turn, modify the reflectance spectrum.
Finally, Figs.~\ref{fig1}(i-l) show the resulting material color corresponding to each reflectance spectrum. This provides a direct visual confirmation that the color of materials is a tunable property controlled by the underlying quantum geometry of its electronic wavefunctions.

\begin{figure}[t]
\includegraphics[width=100mm]{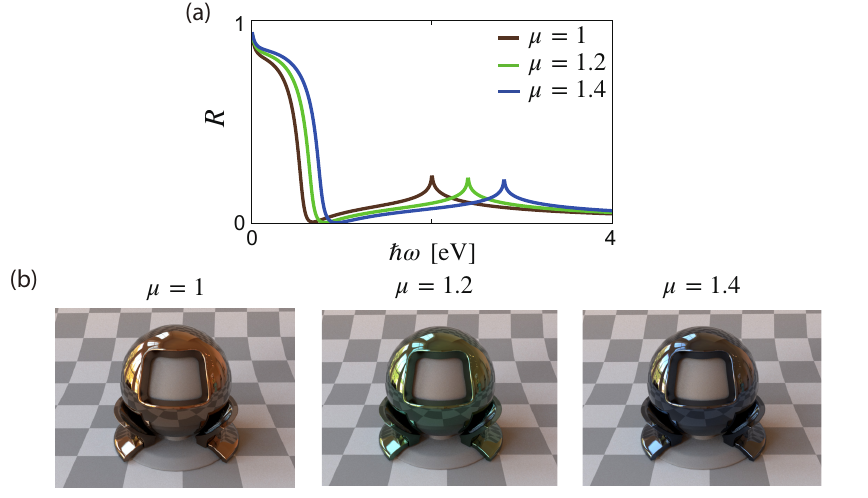} 
\caption{\label{fig2}
(a) Reflectance spectra $R(\omega)$ for different values of the chemical potential, $\mu=1.0$, $1.2$, and $1.4~\text{eV}$. The sharp feature in each spectrum, which corresponds to the interband absorption edge at $\hbar\omega = 2\mu$.
(b) Photorealistic visualizations of the perceived material colors corresponding to the reflectance spectra in (a). The calculations are performed using the parameters $(J_\theta,J_\phi)=(2,2)$, $m=0.1 m_e$, $m_e=0.511~\text{MeV}$, and $\eta=0.1~\text{eV}$.
}
\end{figure}

In addition to the quantum geometry, the chemical potential $\mu$ provides another crucial knob for tuning the color of materials. The value of $\mu$ governs the optical properties through two primary mechanisms. First, it determines the carrier density $n$, which is proportional to $\mu^{3/2}$ in our 3D model. This directly sets the strength of the low-frequency intraband (Drude) response, modifying the metallic character of the reflectance spectrum.
Second, and more dramatically for the perceived color, the chemical potential sets the energy threshold for interband transitions. As dictated by the Heaviside step function $\Theta(\hbar\omega - 2\mu)$ in the optical conductivity, significant interband absorption only occurs when the photon energy $\hbar\omega$ is greater than twice the chemical potential, $2\mu$. This value of $2\mu$ acts as a sharp absorption edge. By varying $\mu$, one can shift this absorption edge across the visible spectrum, thereby controlling which colors are absorbed and which are reflected.
This tunability is illustrated in Fig.~2. Here, we fix the geometric parameters $(J_\theta, J_\phi)=(2,2)$ and plot the reflectance spectra for several values of $\mu$. As shown in Fig.~\ref{fig2}(a), the onset of interband absorption, marked by a sharp feature in the reflectance, systematically shifts to higher frequencies as $\mu$ increases. Consequently, the perceived color of the material undergoes a dramatic transformation, as visualized in Fig.~\ref{fig2}(b).
Therefore, the color of materials is determined by a rich interplay between the quantum geometry and the carrier doping, all set against the backdrop of the fundamental band dispersion of materials. These factors together offer a versatile platform for designing materials with on-demand optical properties.

\subsection{Transparency modulated by quantum geometry}
\subsubsection{Optical Absorbance in 2D Materials}
To model the optical response of a 2D material, we consider the scenario of an electromagnetic wave at normal incidence. The electric and magnetic fields of the wave are oriented parallel to the plane of the 2D sample. The incident power flux, which is the energy passing through a unit area per unit time, is given by the magnitude of the Poynting vector, $S_{\mathrm{inc}} = c\epsilon_0 |\bm{E}|^2$, where $\bm{E}$ is the electric field amplitude in vacuum.

When the wave interacts with the material, the charge carriers are driven by the electric field, leading to Joule heating. The power dissipated per unit area within the 2D material is $P_{\mathrm{abs}} = \mathrm{Re}[\bm{J} \cdot \bm{E}^*] = \mathrm{Re}[\sigma_{2D}(\omega)] |\bm{E}|^2$, where $\sigma_{2D}(\omega)$ is the 2D optical conductivity.

For an atomically thin film, the absorbance $A$—the fraction of incident power absorbed by the material—can be approximated by the ratio of the dissipated power to the incident power:
\begin{align}
    A(\omega) \approx \frac{P_{\mathrm{abs}}}{S_{\mathrm{inc}}} = \frac{\mathrm{Re}[\sigma_{2D}(\omega)]}{c\epsilon_0}. \label{eq:absorbance_2D}
\end{align}
This simple and intuitive relation is valid in the limit of low absorbance ($A \ll 1$), as it approximates the local electric field at the sheet with the incident field, thereby neglecting the back-action from the induced currents. Since energy is conserved, the transmittance $T$, reflectance $R$, and absorbance $A$ must sum to unity: $T+R+A=1$. For many 2D materials where reflectance is also small, the transmittance is directly modulated by the absorbance as $T(\omega) \approx 1 - A(\omega)$. For instance, monolayer and bilayer graphenes are good examples\,\cite{nair2008fine}. Therefore, calculating the real part of the 2D optical conductivity is key to understanding the transmittance.

\begin{figure}[t]
\includegraphics[width=100mm]{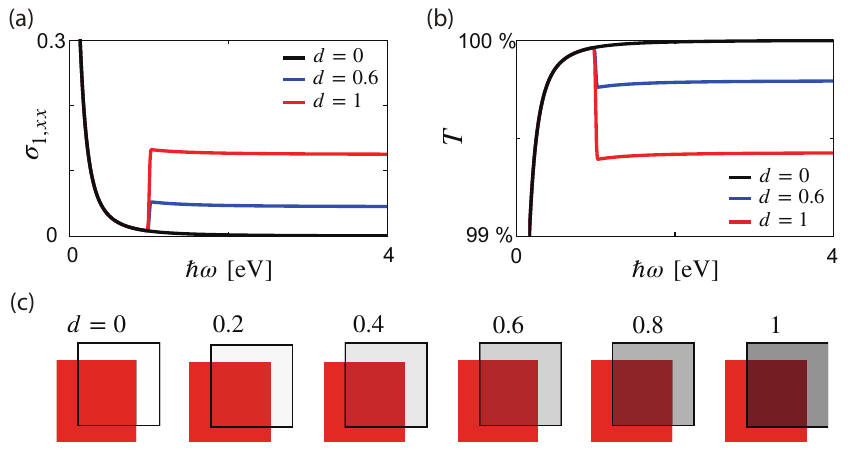} 
\caption{\label{fig3}
(a) Real part of the optical conductivity, $\sigma_{1,xx}(\omega)$, as a function of photon energy $\hbar \omega$ for $d=0$, 0.6, and 1.
(b) Transmittance $T(\hbar\omega)$ of a single 2D sheet for the same values of $d$ shown in (a).
(c) Visualization of the engineered transparency. This panel simulates the appearance of a stack of 50 non-interacting sheets placed over a red background for $d=0$, 0.2, 0.4, 0.6, 0.8, and 1. 
In the calculations, we set $\mu=0.5~\text{eV}$ and $\eta=0.1~\text{eV}$.
}
\end{figure}

\subsubsection{Isotropic Quadratic Band Touching Model}
As a concrete example, we employ an isotropic quadratic band touching model, which allows us to tune the quantum geometry of the system while keeping the band dispersion fixed\,\cite{oh2024thermoelectric,oh2025universal,oh2024revisiting}. For simplicity, we consider a particle-hole symmetric energy dispersion: $\epsilon_\pm(\bm{k}) = \pm\hbar^2 k^2/(2M)$. The Hamiltonian is given by \,\cite{oh2024thermoelectric,oh2025universal,oh2024revisiting}:
\begin{align}
    H(\bm{k}) = \bm{h}(\bm{k}) \cdot \bm{\sigma} , \label{eq:Ham2D}
\end{align}
where the components of $\bm{h}(\bm{k})$ are real quadratic functions of momentum $\bm{k}=(k_x, k_y)$:
\begin{align}
    &h_{x}(\bm{k}) = d \sqrt{1-d^2} \frac{\hbar^2 k_y^2}{M}, \quad
    h_{y}(\bm{k}) = d \frac{\hbar^2 k_x k_y}{M}, \nonumber \\
    &h_{z}(\bm{k}) = \frac{\hbar^2}{2M} \left[k_x^2+(1-2d^2)k_y^2\right],
\end{align}
where $|d|\leq1$.
The dimensionless parameter $d$  encapsulates the quantum geometry of the Bloch wavefunctions. While this parameter significantly alters the inter-band coupling strength, it does not appear in the energy dispersion $\epsilon_\pm(\bm{k})$. In more general contexts, this geometric parameter can be related to the Hilbert-Schmidt quantum distance\,\cite{oh2024thermoelectric,oh2025universal,oh2024revisiting}.

The real part of the diagonal optical conductivity for this model is given by (See Appendix and Ref.\,\cite{oh2025universal} for derivation):
\begin{align}
    \sigma_{1,jj}(\omega) = \underbrace{\frac{e^2}{\hbar}\frac{\mu}{2\pi}\frac{\eta}{(\hbar\omega)^2+\eta^2}}_{\text{Intraband (Drude)}} + \underbrace{\frac{e^2}{8\hbar}d^2 \Theta(\hbar\omega-2\mu)}_{\text{Interband}}. \label{eq:cond_2D_model}
\end{align}
The first term describes the conventional intraband (Drude) response of carriers near the Fermi level $\mu$, broadened by a scattering rate $\eta$. The second term represents the interband response. Notably, the intensity of the interband contribution is independent of the effective mass $M$ and is directly proportional to the square of the geometric parameter, $d^2$. This is a ``mass-invariant universal'' contribution\,\cite{oh2025universal}, which becomes active only when the photon energy $\hbar\omega$ is sufficient to bridge the gap $2\mu$ at the Fermi momentum. This result explicitly demonstrates that the quantum geometry, parameterized by $d$, governs the optical response.

The consequences of this finding are illustrated in Fig.~3. Figure~3(a) plots the real part of the optical conductivity, $\sigma_{1,xx}(\omega)$, for the values of $d=0$, 0.6, and 1. While the low-frequency Drude peak is identical for all cases, the high-frequency interband conductivity appears as a flat plateau whose height is proportional to $d^2$. This directly impacts the optical absorbance $A(\omega)$. Consequently, the transmittance $T(\omega)$ is controlled by the geometric parameter $d$, as shown in Fig.~3(b). 
The transmittance of a single 2D sheet is very high (close to $100\%$), making changes induced by the parameter $d$ difficult to distinguish with the naked eye.
To make this effect visually apparent, we consider a stack of 50 non-interacting sheets, for which the total transmittance would be $T_{\mathrm{total}} = (T(\omega))^{50}$. 
Figure~3(c) illustrates this by simulating the appearance of materials with varying $d$ values (from $0$ to $1$) placed over a red background. As $d$ increases, the material becomes less transparent, causing the red background to appear dimmer. This clearly demonstrates that the transmittance—and thus the perceived transparency—of a material can be engineered via its quantum geometry, even without altering its fundamental energy dispersion.

\section{Conclusion}

In this work, we have demonstrated that quantum geometry can serve as a direct and independent design principle for controlling the intuitive optical properties of materials—their color and transparency. By constructing model systems where the geometry of Bloch states can be tuned while the energy dispersion is held fixed, we have shown that the visual appearance of materials can be a direct manifestation of its underlying quantum geometric structure.

Our investigation of a 3D quadratic band-touching model revealed that the color of materials can be fundamentally geometric in origin. By engineering the momentum-space textures of wavefunctions via tunable geometric integers $(J_\theta, J_\phi)$, while the underlying energy dispersion remained fixed, we produced a palette of distinct colors. This result showcases a mechanism for color generation, moving beyond traditional methods that rely on tuning band gaps through chemical or structural modifications.
This principle was further solidified by our analysis of a 2D system. There, we illustrated how interband conductivity of purely geometric origin provides a direct and quantitative knob to engineer the transparency of materials. 

The central advance of our work lies in demonstrating this decoupling of quantum geometry from the electronic band structure as an engineering tool. While many studies have linked optical properties to the quantum geometry\,\cite{cook2017design, de2017quantized, bhalla2022resonant,ghosh2024probing, ezawa2024analytic,ahn2022riemannian, chen2025dielectric,ezawa2025quantum}, our work moves beyond describing this correlation. By treating geometry as a free parameter, we establish a direct causal link from the wavefunction texture to a targeted, human-perceptible optical property. This approach introduces a new degree of freedom for material design, establishing ``quantum geometry engineering'' as a practical concept for creating metamaterials with bespoke optical functionalities. This opens the door for future investigations, from identifying real materials that host such geometrically tunable properties to exploring the role of geometric engineering in nonlinear optics or achieving dynamic control of color and transparency with external stimuli.

\bibliography{ref.bib}

\begin{thebibliography}{34}%
\makeatletter
\providecommand \@ifxundefined [1]{%
 \@ifx{#1\undefined}
}%
\providecommand \@ifnum [1]{%
 \ifnum #1\expandafter \@firstoftwo
 \else \expandafter \@secondoftwo
 \fi
}%
\providecommand \@ifx [1]{%
 \ifx #1\expandafter \@firstoftwo
 \else \expandafter \@secondoftwo
 \fi
}%
\providecommand \natexlab [1]{#1}%
\providecommand \enquote  [1]{``#1''}%
\providecommand \bibnamefont  [1]{#1}%
\providecommand \bibfnamefont [1]{#1}%
\providecommand \citenamefont [1]{#1}%
\providecommand \href@noop [0]{\@secondoftwo}%
\providecommand \href [0]{\begingroup \@sanitize@url \@href}%
\providecommand \@href[1]{\@@startlink{#1}\@@href}%
\providecommand \@@href[1]{\endgroup#1\@@endlink}%
\providecommand \@sanitize@url [0]{\catcode `\\12\catcode `\$12\catcode
  `\&12\catcode `\#12\catcode `\^12\catcode `\_12\catcode `\%12\relax}%
\providecommand \@@startlink[1]{}%
\providecommand \@@endlink[0]{}%
\providecommand \url  [0]{\begingroup\@sanitize@url \@url }%
\providecommand \@url [1]{\endgroup\@href {#1}{\urlprefix }}%
\providecommand \urlprefix  [0]{URL }%
\providecommand \Eprint [0]{\href }%
\providecommand \doibase [0]{https://doi.org/}%
\providecommand \selectlanguage [0]{\@gobble}%
\providecommand \bibinfo  [0]{\@secondoftwo}%
\providecommand \bibfield  [0]{\@secondoftwo}%
\providecommand \translation [1]{[#1]}%
\providecommand \BibitemOpen [0]{}%
\providecommand \bibitemStop [0]{}%
\providecommand \bibitemNoStop [0]{.\EOS\space}%
\providecommand \EOS [0]{\spacefactor3000\relax}%
\providecommand \BibitemShut  [1]{\csname bibitem#1\endcsname}%
\let\auto@bib@innerbib\@empty
\bibitem [{\citenamefont {Peotta}\ and\ \citenamefont
  {T{\"o}rm{\"a}}(2015)}]{peotta2015superfluidity}%
  \BibitemOpen
  \bibfield  {author} {\bibinfo {author} {\bibfnamefont {S.}~\bibnamefont
  {Peotta}}\ and\ \bibinfo {author} {\bibfnamefont {P.}~\bibnamefont
  {T{\"o}rm{\"a}}},\ }\bibfield  {title} {\bibinfo {title} {Superfluidity in
  topologically nontrivial flat bands},\ }\href
  {https://www.nature.com/articles/ncomms9944} {\bibfield  {journal} {\bibinfo
  {journal} {Nat. Commun.}\ }\textbf {\bibinfo {volume} {6}},\ \bibinfo {pages}
  {8944} (\bibinfo {year} {2015})}\BibitemShut {NoStop}%
\bibitem [{\citenamefont {T{\"o}rm{\"a}}\ \emph {et~al.}(2022)\citenamefont
  {T{\"o}rm{\"a}}, \citenamefont {Peotta},\ and\ \citenamefont
  {Bernevig}}]{torma2022superconductivity}%
  \BibitemOpen
  \bibfield  {author} {\bibinfo {author} {\bibfnamefont {P.}~\bibnamefont
  {T{\"o}rm{\"a}}}, \bibinfo {author} {\bibfnamefont {S.}~\bibnamefont
  {Peotta}},\ and\ \bibinfo {author} {\bibfnamefont {B.~A.}\ \bibnamefont
  {Bernevig}},\ }\bibfield  {title} {\bibinfo {title} {Superconductivity,
  superfluidity and quantum geometry in twisted multilayer systems},\ }\href
  {https://www.nature.com/articles/s42254-022-00466-y} {\bibfield  {journal}
  {\bibinfo  {journal} {Nat. Rev. Phys.}\ }\textbf {\bibinfo {volume} {4}},\
  \bibinfo {pages} {528} (\bibinfo {year} {2022})}\BibitemShut {NoStop}%
\bibitem [{\citenamefont {Ozawa}\ and\ \citenamefont
  {Mera}(2021)}]{ozawa2021relations}%
  \BibitemOpen
  \bibfield  {author} {\bibinfo {author} {\bibfnamefont {T.}~\bibnamefont
  {Ozawa}}\ and\ \bibinfo {author} {\bibfnamefont {B.}~\bibnamefont {Mera}},\
  }\bibfield  {title} {\bibinfo {title} {Relations between topology and the
  quantum metric for {C}hern insulators},\ }\href
  {https://journals.aps.org/prb/abstract/10.1103/PhysRevB.104.045103}
  {\bibfield  {journal} {\bibinfo  {journal} {Phys. Rev. B}\ }\textbf {\bibinfo
  {volume} {104}},\ \bibinfo {pages} {045103} (\bibinfo {year}
  {2021})}\BibitemShut {NoStop}%
\bibitem [{\citenamefont {Yu}\ \emph {et~al.}(2024)\citenamefont {Yu},
  \citenamefont {Bernevig}, \citenamefont {Queiroz}, \citenamefont {Rossi},
  \citenamefont {T{\"o}rm{\"a}},\ and\ \citenamefont {Yang}}]{yu2024quantum}%
  \BibitemOpen
  \bibfield  {author} {\bibinfo {author} {\bibfnamefont {J.}~\bibnamefont
  {Yu}}, \bibinfo {author} {\bibfnamefont {B.~A.}\ \bibnamefont {Bernevig}},
  \bibinfo {author} {\bibfnamefont {R.}~\bibnamefont {Queiroz}}, \bibinfo
  {author} {\bibfnamefont {E.}~\bibnamefont {Rossi}}, \bibinfo {author}
  {\bibfnamefont {P.}~\bibnamefont {T{\"o}rm{\"a}}},\ and\ \bibinfo {author}
  {\bibfnamefont {B.-J.}\ \bibnamefont {Yang}},\ }\bibfield  {title} {\bibinfo
  {title} {Quantum geometry in quantum materials},\ }\href
  {https://doi.org/10.48550/arXiv.2501.00098} {\bibfield  {journal} {\bibinfo
  {journal} {arXiv preprint arXiv:2501.00098}\ } (\bibinfo {year}
  {2024})}\BibitemShut {NoStop}%
\bibitem [{\citenamefont {T{\"o}rm{\"a}}(2023)}]{torma2023essay}%
  \BibitemOpen
  \bibfield  {author} {\bibinfo {author} {\bibfnamefont {P.}~\bibnamefont
  {T{\"o}rm{\"a}}},\ }\bibfield  {title} {\bibinfo {title} {Essay: Where can
  quantum geometry lead us?},\ }\href
  {https://journals.aps.org/prl/abstract/10.1103/PhysRevLett.131.240001}
  {\bibfield  {journal} {\bibinfo  {journal} {Phys. Rev. Lett.}\ }\textbf
  {\bibinfo {volume} {131}},\ \bibinfo {pages} {240001} (\bibinfo {year}
  {2023})}\BibitemShut {NoStop}%
\bibitem [{\citenamefont {Provost}\ and\ \citenamefont
  {Vallee}(1980)}]{provost1980riemannian}%
  \BibitemOpen
  \bibfield  {author} {\bibinfo {author} {\bibfnamefont {J.}~\bibnamefont
  {Provost}}\ and\ \bibinfo {author} {\bibfnamefont {G.}~\bibnamefont
  {Vallee}},\ }\bibfield  {title} {\bibinfo {title} {Riemannian structure on
  manifolds of quantum states},\ }\href
  {https://link.springer.com/article/10.1007/BF02193559} {\bibfield  {journal}
  {\bibinfo  {journal} {Commun. Math. Phys.}\ }\textbf {\bibinfo {volume}
  {76}},\ \bibinfo {pages} {289} (\bibinfo {year} {1980})}\BibitemShut
  {NoStop}%
\bibitem [{\citenamefont {Ma}\ \emph {et~al.}(2010)\citenamefont {Ma},
  \citenamefont {Chen}, \citenamefont {Fan},\ and\ \citenamefont
  {Liu}}]{ma2010abelian}%
  \BibitemOpen
  \bibfield  {author} {\bibinfo {author} {\bibfnamefont {Y.-Q.}\ \bibnamefont
  {Ma}}, \bibinfo {author} {\bibfnamefont {S.}~\bibnamefont {Chen}}, \bibinfo
  {author} {\bibfnamefont {H.}~\bibnamefont {Fan}},\ and\ \bibinfo {author}
  {\bibfnamefont {W.-M.}\ \bibnamefont {Liu}},\ }\bibfield  {title} {\bibinfo
  {title} {Abelian and non-{A}belian quantum geometric tensor},\ }\href
  {https://journals.aps.org/prb/abstract/10.1103/PhysRevB.81.245129} {\bibfield
   {journal} {\bibinfo  {journal} {Phys. Rev. B}\ }\textbf {\bibinfo {volume}
  {81}},\ \bibinfo {pages} {245129} (\bibinfo {year} {2010})}\BibitemShut
  {NoStop}%
\bibitem [{\citenamefont {Berry}(1989)}]{berry1989quantum}%
  \BibitemOpen
  \bibfield  {author} {\bibinfo {author} {\bibfnamefont {M.~V.}\ \bibnamefont
  {Berry}},\ }\bibfield  {title} {\bibinfo {title} {The quantum phase, five
  years after},\ }in\ \href@noop {} {\emph {\bibinfo {booktitle} {Geometric
  Phases in Physics}}},\ \bibinfo {editor} {edited by\ \bibinfo {editor}
  {\bibfnamefont {F.}~\bibnamefont {Wilczek}}\ and\ \bibinfo {editor}
  {\bibfnamefont {E.}~\bibnamefont {Shapere}}}\ (\bibinfo {year} {1989})\ pp.\
  \bibinfo {pages} {3--28}\BibitemShut {NoStop}%
\bibitem [{\citenamefont {Nagaosa}\ \emph {et~al.}(2010)\citenamefont
  {Nagaosa}, \citenamefont {Sinova}, \citenamefont {Onoda}, \citenamefont
  {MacDonald},\ and\ \citenamefont {Ong}}]{nagaosa2010anomalous}%
  \BibitemOpen
  \bibfield  {author} {\bibinfo {author} {\bibfnamefont {N.}~\bibnamefont
  {Nagaosa}}, \bibinfo {author} {\bibfnamefont {J.}~\bibnamefont {Sinova}},
  \bibinfo {author} {\bibfnamefont {S.}~\bibnamefont {Onoda}}, \bibinfo
  {author} {\bibfnamefont {A.~H.}\ \bibnamefont {MacDonald}},\ and\ \bibinfo
  {author} {\bibfnamefont {N.~P.}\ \bibnamefont {Ong}},\ }\bibfield  {title}
  {\bibinfo {title} {Anomalous {H}all effect},\ }\href
  {https://doi.org/10.1103/RevModPhys.82.1539} {\bibfield  {journal} {\bibinfo
  {journal} {Rev. Mod. Phys.}\ }\textbf {\bibinfo {volume} {82}},\ \bibinfo
  {pages} {1539} (\bibinfo {year} {2010})}\BibitemShut {NoStop}%
\bibitem [{\citenamefont {Xiao}\ \emph {et~al.}(2010)\citenamefont {Xiao},
  \citenamefont {Chang},\ and\ \citenamefont {Niu}}]{Xiao_2010_RMP}%
  \BibitemOpen
  \bibfield  {author} {\bibinfo {author} {\bibfnamefont {D.}~\bibnamefont
  {Xiao}}, \bibinfo {author} {\bibfnamefont {M.-C.}\ \bibnamefont {Chang}},\
  and\ \bibinfo {author} {\bibfnamefont {Q.}~\bibnamefont {Niu}},\ }\bibfield
  {title} {\bibinfo {title} {Berry phase effects on electronic properties},\
  }\href {https://doi.org/10.1103/RevModPhys.82.1959} {\bibfield  {journal}
  {\bibinfo  {journal} {Rev. Mod. Phys.}\ }\textbf {\bibinfo {volume} {82}},\
  \bibinfo {pages} {1959} (\bibinfo {year} {2010})}\BibitemShut {NoStop}%
\bibitem [{\citenamefont {Oh}\ \emph {et~al.}(2024{\natexlab{a}})\citenamefont
  {Oh}, \citenamefont {Kim},\ and\ \citenamefont
  {Rhim}}]{oh2024thermoelectric}%
  \BibitemOpen
  \bibfield  {author} {\bibinfo {author} {\bibfnamefont {C.-g.}\ \bibnamefont
  {Oh}}, \bibinfo {author} {\bibfnamefont {K.~W.}\ \bibnamefont {Kim}},\ and\
  \bibinfo {author} {\bibfnamefont {J.-W.}\ \bibnamefont {Rhim}},\ }\bibfield
  {title} {\bibinfo {title} {Thermoelectric transport driven by the
  {H}ilbert--{S}chmidt distance},\ }\href
  {https://advanced.onlinelibrary.wiley.com/doi/full/10.1002/advs.202411313}
  {\bibfield  {journal} {\bibinfo  {journal} {Adv. Sci.}\ }\textbf {\bibinfo
  {volume} {11}},\ \bibinfo {pages} {2411313} (\bibinfo {year}
  {2024}{\natexlab{a}})}\BibitemShut {NoStop}%
\bibitem [{\citenamefont {Liang}\ \emph {et~al.}(2017)\citenamefont {Liang},
  \citenamefont {Vanhala}, \citenamefont {Peotta}, \citenamefont {Siro},
  \citenamefont {Harju},\ and\ \citenamefont {T{\"o}rm{\"a}}}]{liang2017band}%
  \BibitemOpen
  \bibfield  {author} {\bibinfo {author} {\bibfnamefont {L.}~\bibnamefont
  {Liang}}, \bibinfo {author} {\bibfnamefont {T.~I.}\ \bibnamefont {Vanhala}},
  \bibinfo {author} {\bibfnamefont {S.}~\bibnamefont {Peotta}}, \bibinfo
  {author} {\bibfnamefont {T.}~\bibnamefont {Siro}}, \bibinfo {author}
  {\bibfnamefont {A.}~\bibnamefont {Harju}},\ and\ \bibinfo {author}
  {\bibfnamefont {P.}~\bibnamefont {T{\"o}rm{\"a}}},\ }\bibfield  {title}
  {\bibinfo {title} {Band geometry, {B}erry curvature, and superfluid weight},\
  }\href {https://journals.aps.org/prb/abstract/10.1103/PhysRevB.95.024515}
  {\bibfield  {journal} {\bibinfo  {journal} {Phys. Rev. B}\ }\textbf {\bibinfo
  {volume} {95}},\ \bibinfo {pages} {024515} (\bibinfo {year}
  {2017})}\BibitemShut {NoStop}%
\bibitem [{\citenamefont {Rhim}\ \emph {et~al.}(2020)\citenamefont {Rhim},
  \citenamefont {Kim},\ and\ \citenamefont {Yang}}]{rhim2020quantum}%
  \BibitemOpen
  \bibfield  {author} {\bibinfo {author} {\bibfnamefont {J.-W.}\ \bibnamefont
  {Rhim}}, \bibinfo {author} {\bibfnamefont {K.}~\bibnamefont {Kim}},\ and\
  \bibinfo {author} {\bibfnamefont {B.-J.}\ \bibnamefont {Yang}},\ }\bibfield
  {title} {\bibinfo {title} {Quantum distance and anomalous {L}andau levels of
  flat bands},\ }\href {https://www.nature.com/articles/s41586-020-2540-1}
  {\bibfield  {journal} {\bibinfo  {journal} {Nature}\ }\textbf {\bibinfo
  {volume} {584}},\ \bibinfo {pages} {59} (\bibinfo {year} {2020})}\BibitemShut
  {NoStop}%
\bibitem [{\citenamefont {Hwang}\ \emph {et~al.}(2021)\citenamefont {Hwang},
  \citenamefont {Rhim},\ and\ \citenamefont {Yang}}]{hwang2021geometric}%
  \BibitemOpen
  \bibfield  {author} {\bibinfo {author} {\bibfnamefont {Y.}~\bibnamefont
  {Hwang}}, \bibinfo {author} {\bibfnamefont {J.-W.}\ \bibnamefont {Rhim}},\
  and\ \bibinfo {author} {\bibfnamefont {B.-J.}\ \bibnamefont {Yang}},\
  }\bibfield  {title} {\bibinfo {title} {Geometric characterization of
  anomalous {L}andau levels of isolated flat bands},\ }\href
  {https://www.nature.com/articles/s41467-021-26765-z} {\bibfield  {journal}
  {\bibinfo  {journal} {Nat. Commun.}\ }\textbf {\bibinfo {volume} {12}},\
  \bibinfo {pages} {6433} (\bibinfo {year} {2021})}\BibitemShut {NoStop}%
\bibitem [{\citenamefont {Verma}\ \emph {et~al.}(2024)\citenamefont {Verma},
  \citenamefont {Guerci},\ and\ \citenamefont {Queiroz}}]{verma2024geometric}%
  \BibitemOpen
  \bibfield  {author} {\bibinfo {author} {\bibfnamefont {N.}~\bibnamefont
  {Verma}}, \bibinfo {author} {\bibfnamefont {D.}~\bibnamefont {Guerci}},\ and\
  \bibinfo {author} {\bibfnamefont {R.}~\bibnamefont {Queiroz}},\ }\bibfield
  {title} {\bibinfo {title} {Geometric stiffness in interlayer exciton
  condensates},\ }\href {https://doi.org/10.1103/PhysRevLett.132.236001}
  {\bibfield  {journal} {\bibinfo  {journal} {Phys. Rev. Lett.}\ }\textbf
  {\bibinfo {volume} {132}},\ \bibinfo {pages} {236001} (\bibinfo {year}
  {2024})}\BibitemShut {NoStop}%
\bibitem [{\citenamefont {Hu}\ \emph {et~al.}(2022)\citenamefont {Hu},
  \citenamefont {Hyart}, \citenamefont {Pikulin},\ and\ \citenamefont
  {Rossi}}]{HuPRB2022}%
  \BibitemOpen
  \bibfield  {author} {\bibinfo {author} {\bibfnamefont {X.}~\bibnamefont
  {Hu}}, \bibinfo {author} {\bibfnamefont {T.}~\bibnamefont {Hyart}}, \bibinfo
  {author} {\bibfnamefont {D.~I.}\ \bibnamefont {Pikulin}},\ and\ \bibinfo
  {author} {\bibfnamefont {E.}~\bibnamefont {Rossi}},\ }\bibfield  {title}
  {\bibinfo {title} {Quantum-metric-enabled exciton condensate in double
  twisted bilayer graphene},\ }\href
  {https://doi.org/10.1103/PhysRevB.105.L140506} {\bibfield  {journal}
  {\bibinfo  {journal} {Phys. Rev. B}\ }\textbf {\bibinfo {volume} {105}},\
  \bibinfo {pages} {L140506} (\bibinfo {year} {2022})}\BibitemShut {NoStop}%
\bibitem [{\citenamefont {Oh}\ \emph {et~al.}(2024{\natexlab{b}})\citenamefont
  {Oh}, \citenamefont {Rhim},\ and\ \citenamefont {Yang}}]{oh2024revisiting}%
  \BibitemOpen
  \bibfield  {author} {\bibinfo {author} {\bibfnamefont {C.-g.}\ \bibnamefont
  {Oh}}, \bibinfo {author} {\bibfnamefont {J.-W.}\ \bibnamefont {Rhim}},\ and\
  \bibinfo {author} {\bibfnamefont {B.-J.}\ \bibnamefont {Yang}},\ }\bibfield
  {title} {\bibinfo {title} {Revisiting the magnetic responses of bilayer
  graphene from the perspective of quantum distance},\ }\href
  {https://journals.aps.org/prb/abstract/10.1103/PhysRevB.110.155412}
  {\bibfield  {journal} {\bibinfo  {journal} {Phys. Rev. B}\ }\textbf {\bibinfo
  {volume} {110}},\ \bibinfo {pages} {155412} (\bibinfo {year}
  {2024}{\natexlab{b}})}\BibitemShut {NoStop}%
\bibitem [{\citenamefont {Panahiyan}\ \emph {et~al.}(2020)\citenamefont
  {Panahiyan}, \citenamefont {Chen},\ and\ \citenamefont
  {Fritzsche}}]{panahiyan2020fidelity}%
  \BibitemOpen
  \bibfield  {author} {\bibinfo {author} {\bibfnamefont {S.}~\bibnamefont
  {Panahiyan}}, \bibinfo {author} {\bibfnamefont {W.}~\bibnamefont {Chen}},\
  and\ \bibinfo {author} {\bibfnamefont {S.}~\bibnamefont {Fritzsche}},\
  }\bibfield  {title} {\bibinfo {title} {Fidelity susceptibility near
  topological phase transitions in quantum walks},\ }\href
  {https://doi.org/10.1103/PhysRevB.102.134111} {\bibfield  {journal} {\bibinfo
   {journal} {Physical Review B}\ }\textbf {\bibinfo {volume} {102}},\ \bibinfo
  {pages} {134111} (\bibinfo {year} {2020})}\BibitemShut {NoStop}%
\bibitem [{\citenamefont {Oh}\ \emph {et~al.}(2022)\citenamefont {Oh},
  \citenamefont {Cho}, \citenamefont {Park},\ and\ \citenamefont
  {Rhim}}]{oh2022bulk}%
  \BibitemOpen
  \bibfield  {author} {\bibinfo {author} {\bibfnamefont {C.-g.}\ \bibnamefont
  {Oh}}, \bibinfo {author} {\bibfnamefont {D.}~\bibnamefont {Cho}}, \bibinfo
  {author} {\bibfnamefont {S.~Y.}\ \bibnamefont {Park}},\ and\ \bibinfo
  {author} {\bibfnamefont {J.-W.}\ \bibnamefont {Rhim}},\ }\bibfield  {title}
  {\bibinfo {title} {Bulk-interface correspondence from quantum distance in
  flat band systems},\ }\href
  {https://www.nature.com/articles/s42005-022-01102-y} {\bibfield  {journal}
  {\bibinfo  {journal} {Commun. Phys.}\ }\textbf {\bibinfo {volume} {5}},\
  \bibinfo {pages} {320} (\bibinfo {year} {2022})}\BibitemShut {NoStop}%
\bibitem [{\citenamefont {Kim}\ \emph {et~al.}(2023{\natexlab{a}})\citenamefont
  {Kim}, \citenamefont {Oh},\ and\ \citenamefont {Rhim}}]{kim2023general}%
  \BibitemOpen
  \bibfield  {author} {\bibinfo {author} {\bibfnamefont {H.}~\bibnamefont
  {Kim}}, \bibinfo {author} {\bibfnamefont {C.-g.}\ \bibnamefont {Oh}},\ and\
  \bibinfo {author} {\bibfnamefont {J.-W.}\ \bibnamefont {Rhim}},\ }\bibfield
  {title} {\bibinfo {title} {General construction scheme for geometrically
  nontrivial flat band models},\ }\href
  {https://www.nature.com/articles/s42005-023-01407-6} {\bibfield  {journal}
  {\bibinfo  {journal} {Commun. Phys.}\ }\textbf {\bibinfo {volume} {6}},\
  \bibinfo {pages} {305} (\bibinfo {year} {2023}{\natexlab{a}})}\BibitemShut
  {NoStop}%
\bibitem [{\citenamefont {Cook}\ \emph {et~al.}(2017)\citenamefont {Cook},
  \citenamefont {M.~Fregoso}, \citenamefont {De~Juan}, \citenamefont {Coh},\
  and\ \citenamefont {Moore}}]{cook2017design}%
  \BibitemOpen
  \bibfield  {author} {\bibinfo {author} {\bibfnamefont {A.~M.}\ \bibnamefont
  {Cook}}, \bibinfo {author} {\bibfnamefont {B.}~\bibnamefont {M.~Fregoso}},
  \bibinfo {author} {\bibfnamefont {F.}~\bibnamefont {De~Juan}}, \bibinfo
  {author} {\bibfnamefont {S.}~\bibnamefont {Coh}},\ and\ \bibinfo {author}
  {\bibfnamefont {J.~E.}\ \bibnamefont {Moore}},\ }\bibfield  {title} {\bibinfo
  {title} {Design principles for shift current photovoltaics},\ }\href
  {https://www.nature.com/articles/ncomms14176} {\bibfield  {journal} {\bibinfo
   {journal} {Nat. Commun.}\ }\textbf {\bibinfo {volume} {8}},\ \bibinfo
  {pages} {14176} (\bibinfo {year} {2017})}\BibitemShut {NoStop}%
\bibitem [{\citenamefont {De~Juan}\ \emph {et~al.}(2017)\citenamefont
  {De~Juan}, \citenamefont {Grushin}, \citenamefont {Morimoto},\ and\
  \citenamefont {Moore}}]{de2017quantized}%
  \BibitemOpen
  \bibfield  {author} {\bibinfo {author} {\bibfnamefont {F.}~\bibnamefont
  {De~Juan}}, \bibinfo {author} {\bibfnamefont {A.~G.}\ \bibnamefont
  {Grushin}}, \bibinfo {author} {\bibfnamefont {T.}~\bibnamefont {Morimoto}},\
  and\ \bibinfo {author} {\bibfnamefont {J.~E.}\ \bibnamefont {Moore}},\
  }\bibfield  {title} {\bibinfo {title} {Quantized circular photogalvanic
  effect in {W}eyl semimetals},\ }\href
  {https://www.nature.com/articles/ncomms15995} {\bibfield  {journal} {\bibinfo
   {journal} {Nat. Commun.}\ }\textbf {\bibinfo {volume} {8}},\ \bibinfo
  {pages} {15995} (\bibinfo {year} {2017})}\BibitemShut {NoStop}%
\bibitem [{\citenamefont {Bhalla}\ \emph {et~al.}(2022)\citenamefont {Bhalla},
  \citenamefont {Das}, \citenamefont {Culcer},\ and\ \citenamefont
  {Agarwal}}]{bhalla2022resonant}%
  \BibitemOpen
  \bibfield  {author} {\bibinfo {author} {\bibfnamefont {P.}~\bibnamefont
  {Bhalla}}, \bibinfo {author} {\bibfnamefont {K.}~\bibnamefont {Das}},
  \bibinfo {author} {\bibfnamefont {D.}~\bibnamefont {Culcer}},\ and\ \bibinfo
  {author} {\bibfnamefont {A.}~\bibnamefont {Agarwal}},\ }\bibfield  {title}
  {\bibinfo {title} {Resonant second-harmonic generation as a probe of quantum
  geometry},\ }\href
  {https://journals.aps.org/prl/abstract/10.1103/PhysRevLett.129.227401}
  {\bibfield  {journal} {\bibinfo  {journal} {Phys. Rev. Lett.}\ }\textbf
  {\bibinfo {volume} {129}},\ \bibinfo {pages} {227401} (\bibinfo {year}
  {2022})}\BibitemShut {NoStop}%
\bibitem [{\citenamefont {Ghosh}\ \emph {et~al.}(2024)\citenamefont {Ghosh},
  \citenamefont {Onishi}, \citenamefont {Xu}, \citenamefont {Lin},
  \citenamefont {Fu},\ and\ \citenamefont {Bansil}}]{ghosh2024probing}%
  \BibitemOpen
  \bibfield  {author} {\bibinfo {author} {\bibfnamefont {B.}~\bibnamefont
  {Ghosh}}, \bibinfo {author} {\bibfnamefont {Y.}~\bibnamefont {Onishi}},
  \bibinfo {author} {\bibfnamefont {S.-Y.}\ \bibnamefont {Xu}}, \bibinfo
  {author} {\bibfnamefont {H.}~\bibnamefont {Lin}}, \bibinfo {author}
  {\bibfnamefont {L.}~\bibnamefont {Fu}},\ and\ \bibinfo {author}
  {\bibfnamefont {A.}~\bibnamefont {Bansil}},\ }\bibfield  {title} {\bibinfo
  {title} {Probing quantum geometry through optical conductivity and magnetic
  circular dichroism},\ }\href
  {https://www.science.org/doi/10.1126/sciadv.ado1761} {\bibfield  {journal}
  {\bibinfo  {journal} {Sci. Adv.}\ }\textbf {\bibinfo {volume} {10}},\
  \bibinfo {pages} {eado1761} (\bibinfo {year} {2024})}\BibitemShut {NoStop}%
\bibitem [{\citenamefont {Ezawa}(2024)}]{ezawa2024analytic}%
  \BibitemOpen
  \bibfield  {author} {\bibinfo {author} {\bibfnamefont {M.}~\bibnamefont
  {Ezawa}},\ }\bibfield  {title} {\bibinfo {title} {Analytic approach to
  quantum metric and optical conductivity in {D}irac models with parabolic mass
  in arbitrary dimensions},\ }\href
  {https://journals.aps.org/prb/abstract/10.1103/PhysRevB.110.195437}
  {\bibfield  {journal} {\bibinfo  {journal} {Phys. Rev. B}\ }\textbf {\bibinfo
  {volume} {110}},\ \bibinfo {pages} {195437} (\bibinfo {year}
  {2024})}\BibitemShut {NoStop}%
\bibitem [{\citenamefont {Ahn}\ \emph {et~al.}(2022)\citenamefont {Ahn},
  \citenamefont {Guo}, \citenamefont {Nagaosa},\ and\ \citenamefont
  {Vishwanath}}]{ahn2022riemannian}%
  \BibitemOpen
  \bibfield  {author} {\bibinfo {author} {\bibfnamefont {J.}~\bibnamefont
  {Ahn}}, \bibinfo {author} {\bibfnamefont {G.-Y.}\ \bibnamefont {Guo}},
  \bibinfo {author} {\bibfnamefont {N.}~\bibnamefont {Nagaosa}},\ and\ \bibinfo
  {author} {\bibfnamefont {A.}~\bibnamefont {Vishwanath}},\ }\bibfield  {title}
  {\bibinfo {title} {Riemannian geometry of resonant optical responses},\
  }\href {https://www.nature.com/articles/s41567-021-01465-z} {\bibfield
  {journal} {\bibinfo  {journal} {Nat. Phys.}\ }\textbf {\bibinfo {volume}
  {18}},\ \bibinfo {pages} {290} (\bibinfo {year} {2022})}\BibitemShut
  {NoStop}%
\bibitem [{\citenamefont {Chen}(2025)}]{chen2025dielectric}%
  \BibitemOpen
  \bibfield  {author} {\bibinfo {author} {\bibfnamefont {W.}~\bibnamefont
  {Chen}},\ }\bibfield  {title} {\bibinfo {title} {Dielectric and optical
  markers originating from quantum geometry},\ }\href
  {https://doi.org/10.1103/PhysRevB.111.085202} {\bibfield  {journal} {\bibinfo
   {journal} {Physical Review B}\ }\textbf {\bibinfo {volume} {111}},\ \bibinfo
  {pages} {085202} (\bibinfo {year} {2025})}\BibitemShut {NoStop}%
\bibitem [{\citenamefont {Ezawa}(2025)}]{ezawa2025quantum}%
  \BibitemOpen
  \bibfield  {author} {\bibinfo {author} {\bibfnamefont {M.}~\bibnamefont
  {Ezawa}},\ }\bibfield  {title} {\bibinfo {title} {Quantum geometry and
  elliptic optical dichroism in p-wave magnets},\ }\href
  {https://doi.org/10.1103/z9gw-lppc} {\bibfield  {journal} {\bibinfo
  {journal} {Physical Review B}\ }\textbf {\bibinfo {volume} {112}},\ \bibinfo
  {pages} {045302} (\bibinfo {year} {2025})}\BibitemShut {NoStop}%
\bibitem [{\citenamefont {Oh}\ \emph {et~al.}(2025)\citenamefont {Oh},
  \citenamefont {Kim}, \citenamefont {Kim}, \citenamefont {Monserrat},\ and\
  \citenamefont {Rhim}}]{oh2025universal}%
  \BibitemOpen
  \bibfield  {author} {\bibinfo {author} {\bibfnamefont {C.-g.}\ \bibnamefont
  {Oh}}, \bibinfo {author} {\bibfnamefont {S.-W.}\ \bibnamefont {Kim}},
  \bibinfo {author} {\bibfnamefont {K.~W.}\ \bibnamefont {Kim}}, \bibinfo
  {author} {\bibfnamefont {B.}~\bibnamefont {Monserrat}},\ and\ \bibinfo
  {author} {\bibfnamefont {J.-W.}\ \bibnamefont {Rhim}},\ }\bibfield  {title}
  {\bibinfo {title} {Universal optical conductivity from quantum geometry in
  quadratic band-touching semimetals},\ }\href
  {https://doi.org/10.48550/arXiv.2503.18372} {\bibfield  {journal} {\bibinfo
  {journal} {arXiv preprint arXiv:2503.18372}\ } (\bibinfo {year}
  {2025})}\BibitemShut {NoStop}%
\bibitem [{\citenamefont {Nassau}(2001)}]{nassau2001physics}%
  \BibitemOpen
  \bibfield  {author} {\bibinfo {author} {\bibfnamefont {K.}~\bibnamefont
  {Nassau}},\ }\href@noop {} {\emph {\bibinfo {title} {The physics and
  chemistry of color: the fifteen causes of color}}}\ (\bibinfo  {publisher}
  {John Wiley and Sons},\ \bibinfo {year} {2001})\BibitemShut {NoStop}%
\bibitem [{\citenamefont {Fox}(2010)}]{fox2010optical}%
  \BibitemOpen
  \bibfield  {author} {\bibinfo {author} {\bibfnamefont {M.}~\bibnamefont
  {Fox}},\ }\href@noop {} {\emph {\bibinfo {title} {Optical properties of
  solids}}}\ (\bibinfo  {publisher} {Oxford university press},\ \bibinfo {year}
  {2010})\BibitemShut {NoStop}%
\bibitem [{\citenamefont {Kim}\ \emph {et~al.}(2023{\natexlab{b}})\citenamefont
  {Kim}, \citenamefont {Conway}, \citenamefont {Pickard}, \citenamefont
  {Pascut},\ and\ \citenamefont {Monserrat}}]{Kim2023_color}%
  \BibitemOpen
  \bibfield  {author} {\bibinfo {author} {\bibfnamefont {S.-W.}\ \bibnamefont
  {Kim}}, \bibinfo {author} {\bibfnamefont {L.~J.}\ \bibnamefont {Conway}},
  \bibinfo {author} {\bibfnamefont {C.~J.}\ \bibnamefont {Pickard}}, \bibinfo
  {author} {\bibfnamefont {G.~L.}\ \bibnamefont {Pascut}},\ and\ \bibinfo
  {author} {\bibfnamefont {B.}~\bibnamefont {Monserrat}},\ }\bibfield  {title}
  {\bibinfo {title} {Microscopic theory of colour in lutetium hydride},\ }\href
  {https://doi.org/10.1038/s41467-023-42983-z} {\bibfield  {journal} {\bibinfo
  {journal} {Nature Communications}\ }\textbf {\bibinfo {volume} {14}},\
  \bibinfo {pages} {7360} (\bibinfo {year} {2023}{\natexlab{b}})}\BibitemShut
  {NoStop}%
\bibitem [{\citenamefont {Jakob}\ \emph {et~al.}(2022)\citenamefont {Jakob},
  \citenamefont {Speierer}, \citenamefont {Roussel}, \citenamefont
  {Nimier-David}, \citenamefont {Vicini}, \citenamefont {Zeltner},
  \citenamefont {Nicolet}, \citenamefont {Crespo}, \citenamefont {Leroy},\ and\
  \citenamefont {Zhang}}]{mitsuba3}%
  \BibitemOpen
  \bibfield  {author} {\bibinfo {author} {\bibfnamefont {W.}~\bibnamefont
  {Jakob}}, \bibinfo {author} {\bibfnamefont {S.}~\bibnamefont {Speierer}},
  \bibinfo {author} {\bibfnamefont {N.}~\bibnamefont {Roussel}}, \bibinfo
  {author} {\bibfnamefont {M.}~\bibnamefont {Nimier-David}}, \bibinfo {author}
  {\bibfnamefont {D.}~\bibnamefont {Vicini}}, \bibinfo {author} {\bibfnamefont
  {T.}~\bibnamefont {Zeltner}}, \bibinfo {author} {\bibfnamefont
  {B.}~\bibnamefont {Nicolet}}, \bibinfo {author} {\bibfnamefont
  {M.}~\bibnamefont {Crespo}}, \bibinfo {author} {\bibfnamefont
  {V.}~\bibnamefont {Leroy}},\ and\ \bibinfo {author} {\bibfnamefont
  {Z.}~\bibnamefont {Zhang}},\ }\href@noop {} {\bibinfo {title} {Mitsuba 3
  renderer}} (\bibinfo {year} {2022}),\ \bibinfo {note}
  {https://mitsuba-renderer.org}\BibitemShut {NoStop}%
\bibitem [{\citenamefont {Nair}\ \emph {et~al.}(2008)\citenamefont {Nair},
  \citenamefont {Blake}, \citenamefont {Grigorenko}, \citenamefont {Novoselov},
  \citenamefont {Booth}, \citenamefont {Stauber}, \citenamefont {Peres},\ and\
  \citenamefont {Geim}}]{nair2008fine}%
  \BibitemOpen
  \bibfield  {author} {\bibinfo {author} {\bibfnamefont {R.~R.}\ \bibnamefont
  {Nair}}, \bibinfo {author} {\bibfnamefont {P.}~\bibnamefont {Blake}},
  \bibinfo {author} {\bibfnamefont {A.~N.}\ \bibnamefont {Grigorenko}},
  \bibinfo {author} {\bibfnamefont {K.~S.}\ \bibnamefont {Novoselov}}, \bibinfo
  {author} {\bibfnamefont {T.~J.}\ \bibnamefont {Booth}}, \bibinfo {author}
  {\bibfnamefont {T.}~\bibnamefont {Stauber}}, \bibinfo {author} {\bibfnamefont
  {N.~M.}\ \bibnamefont {Peres}},\ and\ \bibinfo {author} {\bibfnamefont
  {A.~K.}\ \bibnamefont {Geim}},\ }\bibfield  {title} {\bibinfo {title} {Fine
  structure constant defines visual transparency of graphene},\ }\href
  {https://www.science.org/doi/10.1126/science.1156965} {\bibfield  {journal}
  {\bibinfo  {journal} {science}\ }\textbf {\bibinfo {volume} {320}},\ \bibinfo
  {pages} {1308} (\bibinfo {year} {2008})}\BibitemShut {NoStop}%
\end{thebibliography}%
\section{Acknowledgments}
\begin{acknowledgments}
The authors thank B. Monserrat for useful discussions. C.O. was supported by Q-STEP, WINGS Program, the University of Tokyo. S.-W.K. was supported by a Leverhulme Trust Early Career Fellowship (ECF-2024-052) and by a UKRI Future Leaders Fellowship [MR/V023926/1].
\end{acknowledgments}

\onecolumngrid

\clearpage

\section{Appendix}

\subsection{Calculation of the optical conductivity}
The linear optical conductivity is given by the Kubo formula:
\begin{align}
\sigma_{ij}(\omega)
= \frac{e^2}{\hbar}
  \int \frac{d^d k}{(2\pi)^d}
  \sum_{n,m}
  F_{nm}(\bm{k})
  \frac{i\,\epsilon_{mn}(\bm{k})\,A^i_{nm}(\bm{k})\,A^j_{mn}(\bm{k})}
       {\epsilon_{nm}(\bm{k}) + \hbar\omega + i\eta}.
\end{align}
Here we specialize to a two‐band model in \(d=2\) or \(3\) dimensions with
\(\epsilon_{\pm}(\bm{k})=\pm\frac{(\hbar k)^2}{2m}\).

\subsection{Intraband contribution}

A straightforward evaluation yields the intraband (Drude) term
\begin{align}
\sigma^{\rm intra}_{ij}(\omega)
= \frac{e^2}{\hbar}
  \int \frac{d^d k}{(2\pi)^d}
  \sum_{n}
  \bigl(-\partial_\epsilon f_n\bigr)\,v^i_n\,v^j_n\,
  \frac{i}{\hbar\omega + i\eta},
\end{align}
where 
\[
v^i_n(\bm{k})
= \frac{1}{\hbar}\frac{\partial \epsilon_n(\bm{k})}{\partial k_i},
\quad
-\partial_\epsilon f_n = \delta\bigl(\epsilon_n - \mu\bigr)
\quad (T=0).
\]
Using the angular integral
\(\displaystyle
 \int d\Omega\,k_i k_j
 = \frac{k^2}{d}\,S_{d-1}\,\delta_{ij},
\)
with \(S_{d-1}=2\pi^{d/2}/\Gamma(d/2)\) and the Fermi momentum \(k_F\) defined by \(\epsilon_+(k_F)=\mu\), one obtains
\begin{align}
\sigma^{\rm intra}_{ij}(\omega)
= \delta_{ij}
  \,\frac{e^2}{\hbar}\,\frac{n}{m}\,
  \frac{i}{\hbar\omega + i\eta},
\qquad
n = \frac{S_{d-1}}{d\,(2\pi)^d}\,k_F^d.
\end{align}

\subsection{Interband contribution of the 3D model}
We now derive the interband conductivity for the model in Eq.~(\ref{eq:Ham3D}) of the main text.
Furthermore, we set the Fermi energy lies in the “+” band and $\omega>0$, so that
\begin{align}
F_{+-}(k)=f_-(k)-f_+(k)
=1-\Theta\!\Bigl(\mu-\tfrac{(\hbar k)^2}{2m}\Bigr)
=\Theta(k-k_F), 
\quad
k_F=\sqrt{\tfrac{2m\mu}{\hbar^2}}.
\end{align}
The Kubo formula becomes
\begin{align}
\sigma^{\rm inter}_{ij}(\omega)
=\frac{e^2}{\hbar}
\int\frac{d^3k}{(2\pi)^3}\,
\Theta(k-k_F)\,
\frac{i\,\epsilon_{+-}(k)\,A^i_{-+}A^j_{+-}}
     {\epsilon_{-+}(k)+\hbar\omega+i\eta},
\end{align}
where \(\epsilon_{+-}(k)=-\epsilon_{-+}(k)=(\hbar k)^2/m\).

We evaluate the angular part of the matrix element $A^i_{+-}A^i_{-+}$, where
\begin{align}
A^i_{+-}(\bm k) = \langle u_+(\bm k) | i\partial_{k_i} | u_-(\bm k) \rangle.
\end{align}
Since 
\begin{align}
\ket{u_+} =
\begin{pmatrix}
\cos\frac{J_\theta\theta}{2}\\[6pt]
e^{iJ_\phi\phi}\sin\frac{J_\theta\theta}{2}
\end{pmatrix},
\qquad
\ket{u_-} =
\begin{pmatrix}
-\,e^{-iJ_\phi\phi}\sin\frac{J_\theta\theta}{2}\\[6pt]
\cos\frac{J_\theta\theta}{2}
\end{pmatrix},
\end{align}

one finds 
\begin{align}
A^i_{-+}A^i_{+-}
=\frac{1}{4}\Bigl[
  J_\theta^2\bigl(\partial_{k_i}\theta\bigr)^2
+ J_\phi^2\sin^2\!\bigl(J_\theta\theta\bigr)\,\bigl(\partial_{k_i}\phi\bigr)^2
\Bigr].
\end{align}

If we take angular integral, one get
   \begin{align}
   \int d\Omega\,A^i_{-+}A^i_{+-}
   =\frac{J_\theta^2 C_\theta + J_\phi^2\,C_\phi}{4k^2},
   \end{align}
where
\begin{align}
    &C^i_\theta = \frac{\pi(2+6\delta_{iz})}{3}, 
~~~~C^i_\phi(J_\theta) =(1-\delta_{iz}) \pi \int_0^\pi d\theta\frac{\sin^2(J_\theta \theta)}{\sin (\theta)}.
\end{align}
If we insert this, we get
\begin{align}
\sigma^{\rm inter}_{jj}(\omega)
=\frac{e^2}{\hbar}
\int\frac{dk}{(2\pi)^3}\,
\Theta(k-k_F)\,
\frac{i\,\epsilon_{+-}(k)}
     {\epsilon_{-+}(k)+\hbar\omega+i\eta}\frac{J_\theta^2 C_\theta^j + J_\phi^2\,C_\phi^j}{4}.
\end{align}
From this, we get 
\begin{align}
&\mathrm{Re}[\sigma^{\rm inter}_{jj}(\omega)]
=\frac{e^2}{\hbar}\frac{J_\theta^2 C_\theta^j + J_\phi^2\,C_\phi^j}{32\pi^3}
\int_{k_F}^{k_c} dk\,
\frac{\epsilon_{+-}(k)\eta}
     {(\epsilon_{-+}(k)+\hbar\omega)^2+\eta^2}\\
&\mathrm{Im}[\sigma^{\rm inter}_{jj}(\omega)]
=\frac{e^2}{\hbar}\frac{J_\theta^2 C_\theta^j + J_\phi^2\,C_\phi^j}{32\pi^3}
\int_{k_F}^{k_c} dk\,
\frac{\epsilon_{+-}(k)(\epsilon_{-+}(k)+\hbar\omega)}
     {(\epsilon_{-+}(k)+\hbar\omega)^2+\eta^2},
\end{align}
where $k_c$ is a cutoff momentum.
For small $\eta$, we get
\begin{align}
\mathrm{Re}[\sigma^{\rm inter}_{jj}(\omega)]
&=\frac{e^2}{\hbar}\frac{\sqrt{m\omega/\hbar}}{32\pi^2}\frac{J_\theta^2 C_\theta^j + J_\phi^2\,C_\phi^j}{2}\Theta(\hbar\omega-2\mu),\\
\mathrm{Im}[\sigma^{\rm inter}_{jj}(\omega)]
&=-\frac{e^2}{\hbar}\frac{J_\theta^2 C_\theta^j + J_\phi^2\,C_\phi^j}{32\pi^2}\frac{\sqrt{m\omega/\hbar}}{2}\ln \bigg|\frac{(k_c-\sqrt{m\omega/\hbar})(k_F+\sqrt{m\omega/\hbar})}{(k_c+\sqrt{m\omega/\hbar})(k_F-\sqrt{m\omega/\hbar})}\bigg|\\
&\approx -\frac{e^2}{\hbar}\frac{J_\theta^2 C_\theta^j + J_\phi^2\,C_\phi^j}{32\pi^2}\frac{\sqrt{m\omega/\hbar}}{2}\ln \bigg|\frac{(k_F+\sqrt{m\omega/\hbar})}{(k_F-\sqrt{m\omega/\hbar})}\bigg|,
\end{align}
where $(\hbar k_F)^2/(2m) =\mu$
In this calculation, we drop the $\omega$-independent background constant. Furthermore, one can verify that the Kramers-Kronig relation, Eq.~(\ref{eq:KK_rel}), is satisfied.
\end{document}